\documentclass[prl,preprint,amsmath,amssymb,showpacs,showkeys,superscriptaddress]{revtex4}
\usepackage{graphicx}
\usepackage{dcolumn}
\usepackage{hyperref}
\newcommand{\be}{\begin{equation}}
\newcommand{\ee}{\end{equation}}
\newcommand{\ba}{\begin{eqnarray}}
\newcommand{\ea}{\end{eqnarray}}

\begin{document}

\title{Virial coefficients and demixing of athermal nonadditive mixtures}
\author{G. Pellicane}
\altaffiliation{Corresponding Author} \email{\tt
giuseppe.pellicane@unime.it} \affiliation{Universit\`a degli Studi
di Messina, Dipartimento di Fisica \\ Contrada Papardo, 98166
Messina, Italy}
\author{F. Saija}
\email{\tt saija@me.cnr.it}
\affiliation{CNR - Istituto per i Processi Chimico-Fisici,
Sezione di Messina,\\Via La Farina 237, 98123 Messina, Italy}
\author{C. Caccamo}
\email{\tt carlo.caccamo@unime.it} \affiliation{Universit\`a degli
Studi di Messina, Dipartimento di Fisica \\ Contrada Papardo,
98166 Messina, Italy}
\author{P. V. Giaquinta}
\email{\tt paolo.giaquinta@unime.it} \affiliation{Universit\`a
degli Studi di Messina, Dipartimento di Fisica \\ Contrada
Papardo, 98166 Messina, Italy}
\date{\today}

\begin{abstract}
We compute the fourth virial coefficient of a binary nonadditive
hard-sphere mixture over a wide range of deviations from diameter
additivity and size ratios. Hinging on this knowledge, we build up a
$y$ expansion [B.~Barboy and W.~N.~Gelbart, J. Chem. Phys. {\bf 71},
3053 (1979)] in order to trace the fluid-fluid coexistence lines
which we then compare with the available Gibbs-ensemble Monte Carlo
data and with the estimates obtained through two refined
integral-equation theories of the fluid state. We find that in a
regime of moderately negative nonadditivity and largely asymmetric
diameters, relevant to the modelling of sterically and
electrostatically stabilized colloidal mixtures, the fluid-fluid
critical point is unstable with respect to crystallization.
\end{abstract}
\maketitle
\section{Introduction}
A binary mixture of hard spheres is an established prototype model
for understanding excluded-volume phenomena relevant to the
equilibrium statistical mechanics of some complex fluids
\cite{bou-nez86}. The model is characterized by the impenetrable
diameters of the two species $\sigma_{1}$ and $\sigma_{2}$ and by a
cross diameter $\sigma_{12} = (1/2)(\sigma_{1} + \sigma_{2})(1 +
\Delta)$, where the dimensionless quantity $\Delta$ accounts for
deviations of the inter-species repulsive interactions from
additivity. As is well known, in a real system the repulsive effects
are associated with the overlap between the electron clouds of
colliding molecules. However, even in a rare-gas mixture the Lorentz
rule ($\Delta = 0$) is not systematically satisfied
~\cite{rowlinson}. Deviations from additivity are expected to become
more and more pronounced with increasing pressures. In this respect,
notwithstanding its crudeness, the hard-sphere model contains some
basic features which can be used to sketch, for example, the phase
behavior of binary fluids at very high pressures ($\sim 1$ GPa)
\cite{shout}. Deviations from additivity are also manifest in the
``effective'' interactions which characterize some complex
polydisperse fluids such as colloid-polymer mixtures \cite{louis}.

In general, a bidisperse mixture of hard spheres may exist in a
surprisingly rich variety of both ordered and disordered phases
whose formation is driven by entropic effects which depend on the
diameter ratio $q=\sigma_{2}/\sigma_{1}$ (we conventionally identify
the larger particles as species $1$ in such a way that $q\leq1$), on
the deviation from additivity $\Delta$, and on the mole fractions
$x_1$ and $x_2=1-x_1$ of the two species, where $x_1 =
N_1/(N_1+N_2)$ \cite{frenk}. For largely asymmetric size ratios ($q
< 0.3$) and in the absence of nonadditive effects ($\Delta=0$), a
mixture tends to phase separate, as a result of osmotic depletion
effects, when the larger particles are sufficiently dilute
\cite{bh1}. However, Monte Carlo (MC) simulations show that in these
conditions the demixing transition is preempted by the freezing of
the larger particles~\cite{dre}.

The thermodynamic and phase-stability properties of a system turn
out to be considerably affected by an even modest degree of
nonadditivity. In a nonadditive hard-sphere (NAHS) mixture, the
dominant mechanism underlying phase separation is different than in
the additive case. First of all, the positive or negative nature of
the nonadditivity is critical as to the type of thermodynamic
behavior that is exhibited by the mixture. For $\Delta < 0$ the
model shows a trend towards hetero-coordination and is able to
reproduce the large compositional fluctuations that are observed in
some liquid and amorphous mixtures \cite{albas,gaz-past1,gaz-past2}.
On the contrary, for $\Delta > 0$ even an equal-sized NAHS mixture
exhibits phase separation for high enough densities. This phenomenon
is due to the extra repulsion between unlike spheres, which promotes
the homo-coordination. The symmetric (equal-diameter) case has been
investigated with a variety of theoretical methods
\cite{tenne,mazo,nix,bpgg,gazzillo,lala,kahl,sfg,spg}. A number of
phenomenological equations of state have been proposed as well
\cite{ree2,hamad1,hamad2}. Furthermore, computer simulation studies
have been performed for both negative and positive values of the
nonadditivity parameter \cite{melynk,adams,gapa,amar}.
Correspondingly, the critical density as well as the universality
class which the NAHS model belongs to have been determined
\cite{yethi,goz}.

So far, the asymmetric case has been studied much less than the
symmetric one. The first simulation experiment of a NAHS mixture
with similar diameters of the two species was carried out by Rovere
and Pastore who traced the phase diagram by means of the
Gibbs-ensemble Monte Carlo (GEMC) method~\cite{rovere}. Hamad
performed Molecular Dynamics (MD) simulations and calculated the
compressibility factor of a NAHS mixture for a number of values of
$q$ and $\Delta$~\cite{hamad3}. GEMC simulations of a highly
asymmetric ($q = 0.1$) binary mixture show that fluid-fluid phase
separation may occur on account of a rather small amount of
nonadditivity~\cite{dij2}, as had been already conjectured by Biben
and Hansen~\cite{bh2}. A reliable theoretical approach to study the
phase diagram of NAHS mixture in a strongly asymmetric regime is
based on the calculation of the effective potential between the
larger particles. Such an effective potential is obtained upon
integrating out the degrees of freedom associated with the smaller
spheres \cite{lou2,re,lr,louis1}. Recently, we have shown, by means
of integral-equation theories, that a stable fluid-fluid phase
separation may occur for moderate size ratios and for a small amount
of nonadditivity~\cite{pellican06}. Another useful theoretical
approach is based on the calculation of the residual multiparticle
entropy complemented by MC numerical simulations~\cite{sg}. This
approach has been applied by Saija and Giaquinta to study the weakly
asymmetric case ($q \geq 0.75$).

Some efforts in the direction of studying NAHS mixtures by means of
analytical theories have been also made by developing more refined
equations of state (EOS) as compared with those obtained via a crude
virial expansion. Schaink introduced an EOS whose derivation is
reminiscent of the scaled particle theory and is valid in the
symmetric as well as in the asymmetric case~\cite{schaink}. More
recently, Santos and coworkers proposed an EOS that relies on the
exact second and third virial coefficients and requires as an input
the compressibility factor of the one-component
system~\cite{santos}.

In this paper we evaluate the fourth virial coefficient of a
bidisperse NAHS mixture for a number of diameter ratios which span a
wide range of accessible values and for $-0.05 \le \Delta \le 0.5$.
As far as we know, the only published virial coefficients beyond the
third one are due to Vlasov and Masters~\cite{vlasov}. However,
these authors computed the virial coefficients of the mixture up to
the sixth order for the size ratio $q = 0.1$ only and for just one
positive value of the nonadditivity $(\Delta = 0.1)$.

Even a reduced number of virial coefficients can be used to
construct approximate expansions which turn out to be quite reliable
over a wider density range than the original virial
expansion~\cite{bou-nez86}. It is in this conceptual framework that
we aim at probing in this paper the reliability of the so-called
{\it y} expansion, originally proposed by Barboy and Gelbart
(BG)~\cite{barboy1}. We use this approach to calculate the
fluid-fluid phase separation lines and compare the results with the
available Gibbs-ensemble Monte Carlo data and with the results of
some reliable integral-equation closures of the Ornstein-Zernike
equations~\cite{pellican06}.

The paper is organized as follows: In Section II we present the
numerical procedure used to calculate the fourth virial coefficient
and the explicit expression of the BG EOS for a binary hard-sphere
mixture. In Section III we present the phase diagram calculated for
a number of size ratios. Some concluding remarks are finally given
in Section IV.

\section{The virial coefficients and the Barboy-Gelbart equation of state}

The virial expansion can be written as:
\begin{equation}
\beta P = \rho + B \rho^2 + C \rho^3 + D \rho^4 + \ldots
\label{s:bpr}
\end{equation}
where {\it P} is the pressure, $\beta$ is the inverse temperature in
units of the Boltzmann constant, and $\rho = \rho_1 + \rho_2$,
$\rho_i$ $(i=1,2)$ being the particle number density of the {\it
i}-th species. In a mixture, at variance with the one-component
case, the virial coefficients $B, C, D \ldots$ also depend on the
relative concentration of the two species. In particular, the
fourth-order coefficient reads:
\begin{equation}
D = D_{1111}x_1^4 + 4D_{1112}x_1^3x_2 + 6D_{1122}x_1^2x_2^2 +
    4D_{1222}x_1x_2^3 + D_{2222}x_2^4
\label{s:ddix}
\end{equation}

\noindent In Eq.(\ref{s:ddix}) the coefficients $D_{1111}$ and
$D_{2222}$ can be calculated through the corresponding expression
for a monodisperse fluid of particles with diameter $\sigma_1$ or
$\sigma_2$, respectively. The cluster integrals
$D_{1112}$ and $D_{1122}$ can be represented with the following four-point color graphs: \\
\unitlength=1.0pt
\begin{picture}(50,50)(150,850)
\multiput(250,880)(50,0){4}{\circle {6}}
\multiput(270,880)(50,0){4}{\circle {6}}
\multiput(250,860)(50,0){4}{\circle {6}}
\multiput(270,860)(50,0){4}{\circle*{6}}
\multiput(253,880)(50,0){4}{\line(1,0){14}}
\multiput(253,860)(50,0){4}{\line(1,0){14}}
\multiput(250,863)(50,0){4}{\line(0,1){14}}
\multiput(270,863)(50,0){4}{\line(0,1){14}}
\multiput(302.5,862.5)(100,0){2}{\line(1,1) {15.5}}
\multiput(370,860)(50,0){2} {\line(-1,1) {18}}
\put(205,870) {\makebox(0,0){$D_{1112} = - \frac {1}{8} \Biggl [ \ 3$}}
\put(285,870) {\makebox(0,0){$+\ 3$}}
\put(335,870) {\makebox(0,0){$+\ 3$}}
\put(385,870) {\makebox(0,0){$+ $}}
\put(435,870) {\makebox(0,0){$\Biggr ] $}}
\end{picture}
\vspace{-1.1cm}
\begin{equation}
\label{s:d1112}
\end{equation}
\unitlength=1.0pt
\begin{picture}(50,50)(150,850)
\multiput(235,870)(45,0){6}{\circle*{6}}
\multiput(255,870)(45,0){6}{\circle {6}}
\multiput(235,850)(90,0){2}{\circle*{6}}
\multiput(255,850)(25,0){2}{\circle {6}}
\multiput(300,850)(90,0){2}{\circle*{6}}
\multiput(345,850)(25,0){2}{\circle {6}}
\multiput(415,850)(65,0){2}{\circle {6}}
\multiput(435,850)(25,0){2}{\circle*{6}}
\multiput(238,870)(45,0){6}{\line(1,0){14}}
\multiput(238,850)(45,0){6}{\line(1,0){14}}
\multiput(235,853)(45,0){6}{\line(0,1){14}}
\multiput(255,853)(45,0){6}{\line(0,1){14}}
\put(325,850)     {\line(1,1) {18}}
\put(372.5,852.5) {\line(1,1) {15.5}}
\put(415,870)     {\line(1,-1){18}}
\put(460,850)     {\line(1,1) {18}}
\put(460,870)     {\line(1,-1){17.5}}
\put(195,860) {\makebox(0,0){$D_{1122} = - \frac {1}{8} \Biggl [ {\ }2$}}
\put(267,860) {\makebox(0,0){$ + $}}
\put(312,860) {\makebox(0,0){$+\ 4$}}
\put(357,860) {\makebox(0,0){$+ $}}
\put(402,860) {\makebox(0,0){$+ $}}
\put(447,860) {\makebox(0,0){$+ $}}
\put(490,860) {\makebox(0,0){$\Biggr ] $}}
\end{picture}
\vspace{0.2cm}
\begin{equation}
\label{s:d1122}
\end{equation}
\noindent The open and solid circles identify in each graph
particles belonging to species 1 and 2, respectively. Each bond
contributes a factor to the integrand in the form of a Mayer step
function. Space integration is carried out over all the vertices of
the graph. We refer the reader for more details on the algorithm to~\cite{fiumara}.

The coefficient $D_{1222}$ is obtained from the general expression
of $D_{1112}$ by merely interchanging the larger with the smaller
particles and vice-versa. The Monte Carlo results for the
composition-independent coefficients $D_{\alpha \beta \gamma
\delta}$ (with $\alpha,\beta,\gamma,\delta=1,2$) are given in tables
\ref{t1}-\ref{t6} over the whole range of $q$ for $\Delta \leq 0.5$.
As $q \rightarrow 1$, all the coefficients tend to the value
$2.6362...$ (in units of $\sigma_{1}^{9}$) of the one-component
fluid. In the opposite limit of $q \rightarrow 0$ the partial
coefficient $D_{1112}$ approaches the exact result $1/4(\pi
\sigma_{1}^{3}/6)$ \cite{fiumara}. The knowledge of the fourth
virial coefficient allows one to derive the BG expansion up to the
fourth order upon expanding the EOS in powers of the variables $y_i
= \rho_i/(1-\eta)$, where $\eta=\pi \rho_1 \sigma_{1}^{3}/6 + \pi
\rho_2 \sigma_{2}^{3}/6$:

\begin{equation}
\label{bg}
\beta P = \sum_\alpha y_\alpha + \sum_{\alpha\beta} A_{\alpha\beta} y_{\alpha}y_{\beta} +
\sum_{\alpha\beta\gamma} A_{\alpha\beta\gamma} y_{\alpha}y_{\beta}y_{\gamma} +
\sum_{\alpha\beta\gamma\delta} A_{\alpha\beta\gamma\delta} y_{\alpha}y_{\beta}y_{\gamma}y_{\delta}
\end{equation}

\noindent where the coefficients
\begin{equation}
A_{\alpha\beta}=B_{\alpha\beta} - \frac{\pi}{6} \frac{\sigma_{\alpha}^{3}+\sigma_{\beta}^{3}}{2}
\end{equation}
\begin{equation}
A_{\alpha\beta\gamma}=C_{\alpha\beta\gamma} -
\frac{\pi}{9} (\sigma_{\gamma}^{3} A_{\alpha\beta} + \sigma_{\alpha}^{3} A_{\gamma\beta} + \sigma_{\beta}^{3} A_{\alpha\gamma}) -
\frac{\pi^2}{108}(\sigma_{\beta}^{3}\sigma_{\gamma}^{3} + \sigma_{\alpha}^{3}\sigma_{\gamma}^{3} + \sigma_{\beta}^{3}\sigma_{\alpha}^{3})
\end{equation}
\begin{eqnarray}
A_{\alpha\beta\gamma\delta} & = & D_{\alpha\beta\gamma\delta}
- \frac{\pi}{8}(\sigma_{\delta}^{3} A_{\alpha\beta\gamma} + \sigma_{\alpha}^{3} A_{\delta\beta\gamma} + \sigma_{\beta}^{3} A_{\alpha\delta\gamma} + \sigma_{\gamma}^{3} A_{\alpha\beta\delta}) \nonumber\\
& - & \frac{\pi^2}{72} (\sigma_{\gamma}^{3}\sigma_{\delta}^{3} A_{\alpha\beta} + \sigma_{\beta}^{3}\sigma_{\delta}^{3}A_{\alpha\gamma} + \sigma_{\alpha}^{3}\sigma_{\beta}^{3}A_{\gamma\beta} + \sigma_{\gamma}^{3}\sigma_{\alpha}^{3}A_{\delta\beta} + \sigma_{\gamma}^{3}\sigma_{\beta}^{3}A_{\alpha\delta} + \sigma_{\alpha}^{3}\sigma_{\beta}^{3}A_{\delta\gamma}) \nonumber\\
& - & \frac{\pi^3}{864}(\sigma_{\beta}^{3}\sigma_{\gamma}^{3}\sigma_{\delta}^{3} + \sigma_{\alpha}^{3}\sigma_{\gamma}^{3}\sigma_{\delta}^{3} + \sigma_{\beta}^{3}\sigma_{\alpha}^{3}\sigma_{\delta}^{3} + \sigma_{\beta}^{3}\sigma_{\gamma}^{3}\sigma_{\alpha}^{3})
\end{eqnarray}

\noindent are determined so as to ensure the low-density expansion
of Eq.(\ref{bg}) to coincide with the fourth-order virial expansion
reported in Eq.(\ref{s:bpr}). The quantities $B_{\alpha\beta}$ and
$C_{\alpha\beta\gamma}$ are the second and third virial
coefficients, respectively, which are available analytically
\cite{bou-nez86}. In the limit $\Delta \rightarrow 0$, the first three
terms of Eq.(\ref{bg}), which correspond to the third order
expansion of the BG EOS, reduce to the compressibility
equation-of-state of the Percus-Yevick theory. Upon rearranging the
above relations, we can write:
\begin{equation}
\beta P=\sum_{i=1}^{\infty} \frac{\rho^i A_i(x)}{(1 - \eta)^{i}}
\end{equation}
where
\begin{eqnarray}
A_1(x) & = & 1                                                         \nonumber\\
A_2(x) & = & x_{1}^{2}A_{11}+2x_{1}(1-x_{1})A_{12}+(1-x_{1})^{2}A_{22} \nonumber\\
A_3(x) & = & x_{1}^{3}A_{111}+3x_{1}^{2}(1-x_{1})A_{112}+3x(1-x_{1})^{2}A_{122}+(1-x_{1})^{3}A_{222} \nonumber\\
A_4(x) & = & x_{1}^{4}A_{1111}+4x_{1}^{3}(1-x_{1})A_{1112}+6x_{1}^{2}(1-x_{1})^{2}A_{1122}+4x_{1}(1-x_{1})^{3}A_{1222} \nonumber\\
& + & (1-x_{1})^{4}A_{2222}\nonumber
\end{eqnarray}

As we shall see in the following Section, the inclusion of the
fourth virial coefficient definitely improves the performance of the
BG EOS over the third-order one. A more convincing test of the
reliability of the BG EOS including the fourth virial coefficient is
provided by the comparison of the predictions for the fluid-fluid
phase separation threshold with the numerical simulation data. In
order to trace the phase-coexistence curve, we need the Gibbs free
energy. We start by calculating the Helmholtz free energy:

\begin{eqnarray}
\frac{{\beta} A}{N} & = & \int_{0}^{\rho} \Big(\frac{Z}
{\rho}\Big)
                      d{\rho} {}
                  \nonumber\\
                      & = & {} [\ln\big({\rho}{\Lambda}^3 \big) - 1] +
                      \sum_{k=1}^{2}x_{k}\ln(x_{k}) -
                      \ln(1 -\eta) + \sum_{k=1}^{\infty} \frac{\rho^i A_{i+1}(x)}{i(1 - \eta)^{i}}
                  \label{free}
\end{eqnarray}
\noindent where $Z=\beta P/ \rho$ is the compressibility factor; the
Gibbs free energy then follows as~\cite{hansen,bijor}:

\begin{equation}
\frac{\beta G}{N} = \frac{\beta A}{N} + Z. \label{gtot}
\end{equation}

The fluid-fluid coexistence curves are obtained through the
common-tangent construction of the Gibbs free energy at constant
pressure.

\section{Results and Discussion}
\begin{figure}
\begin{center}
\includegraphics[width=11cm,angle=0]{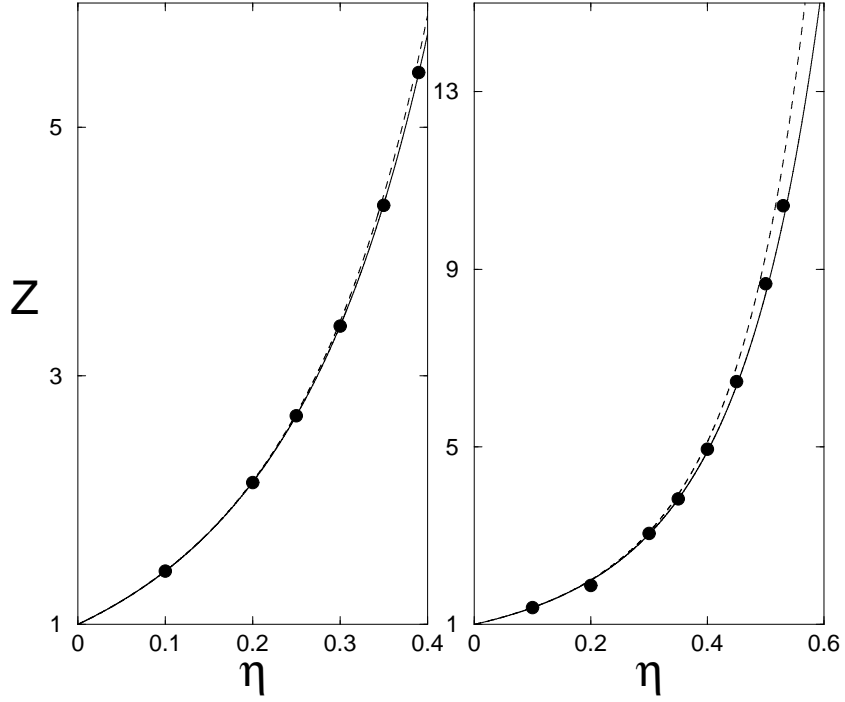}
\end{center}
\caption{Compressibility factor plotted as a function of the total packing
fraction for $q=1/3$, $x_1=0.5$: left panel, $\Delta=-0.05$; right panel, $\Delta=0.05$. The continuous and dashed lines
represent the BG4 and BG3 estimates, respectively. The solid circles
are the MD results by~\cite{hamad3}. The corresponding fourth-order partial virial
coefficients calculated for $\Delta=-0.05$ are: D1112=$0.2304(6)$,
D1122=$2.037(5)\cdot10^{-2}$, D1222=$1.719(1)\cdot10^{-3}$.}
\label{fig1}
\end{figure}

\begin{figure}
\begin{center}
\includegraphics[width=11cm,angle=0]{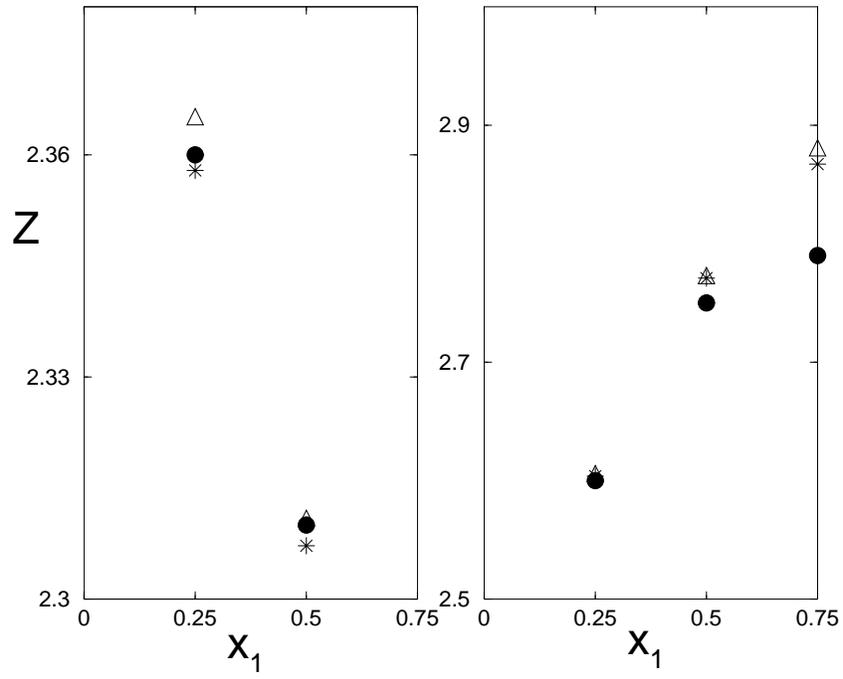}
\end{center}
\caption{Compressibility factor plotted as a function of the mole fraction
for $\Delta = 0.2$: left panel, $q=0.25$; right panel, $q=0.5$. Stars: BG4; triangles: BG3; circles: MD results~\cite{hamad3}.} \label{fig2}
\end{figure}

We present in Figs.~\ref{fig1} and \ref{fig2} the compressibility
factor calculated for different values of the size ratio and of the
nonadditivity parameter. The fourth-order BG expansion (BG4) of the
EOS is in better agreement with the simulation data provided by
Hamad~\cite{hamad3} than the corresponding third-order expansion
(BG3), expecially for large packing fractions. However, as shown in
Fig.~\ref{fig2}, this improvement is not systematic.

\subsection{Species with not too dissimilar sizes}
\begin{figure}
\begin{center}
\includegraphics[width=11cm,angle=0]{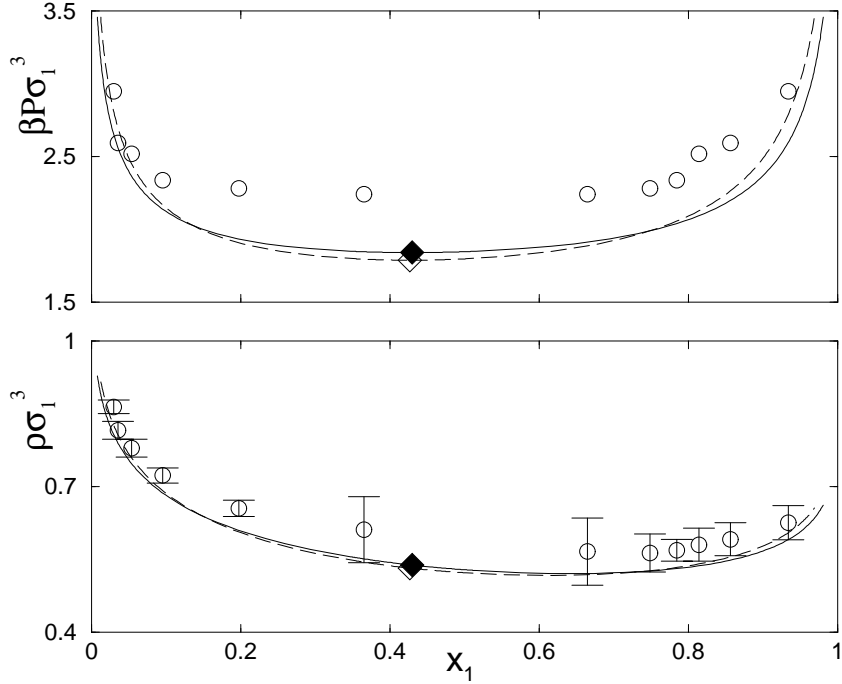}
\end{center}
\caption{Fluid-fluid coexistence in the pressure-composition plane (top panel) and density-composition plane (bottom
panel) plotted for $q=0.8333$ and $\Delta=0.182$. The continuous and dashed curves represent the BG4 and BG3 binodal
lines, respectively. The critical point is identified with solid (BG4) and open (BG3) diamonds. Open circles, with the
corresponding error bars in the density-composition plane, represent the GEMC simulation data~\cite{rovere}. The
corresponding fourth-order partial virial coefficients are: D1112=$3.752(5)$, D1122=$2.457(3)$, D1222=$1.459(1)$. }
\label{fig3}
\end{figure}

We compare in Fig.~\ref{fig3} our results for the phase
coexistence thresholds with those obtained by Rovere and Pastore
for $q=0.8333$ and $\Delta=0.182$ in their GEMC
simulations~\cite{rovere}. Unfortunately, we could not report in
the figure the simulation critical parameters explicitly because
the coexistence curve becomes rather flat around there and a
reasonable determination of them
 would typically require very long computations~\cite{rovere}. As
shown in the top panel of Fig.~\ref{fig3}, both BG estimates of
the EOS appear to underestimate the critical pressure, which is
located at the minimum of the phase coexistence curve, by about
$20\%$. However, on account of the large fluctuations that affect
the system on approaching the critical point, it may be possible
that the pressure measured inside the two simulation boxes is not
sufficiently stable against the value imposed from outside. This
fact might be partially responsible for the discrepancy observed
between the GEMC data and the BG estimates. The growth of
fluctuations near the critical point is apparent in the bottom
panel of Fig.~\ref{fig3} where we report the error bars associated
with the GEMC results for the coexistence densities plotted as a
function of the relative composition. In fact, in the
density-concentration plane, the present theoretical predictions
fall inside the simulation error bars around the critical point.
Overall, the BG4 critical pressure and density show a better
agreement with the GEMC simulation results as compared with their
BG3 counterparts.

\begin{figure}
\begin{center}
\includegraphics[width=11cm,angle=0]{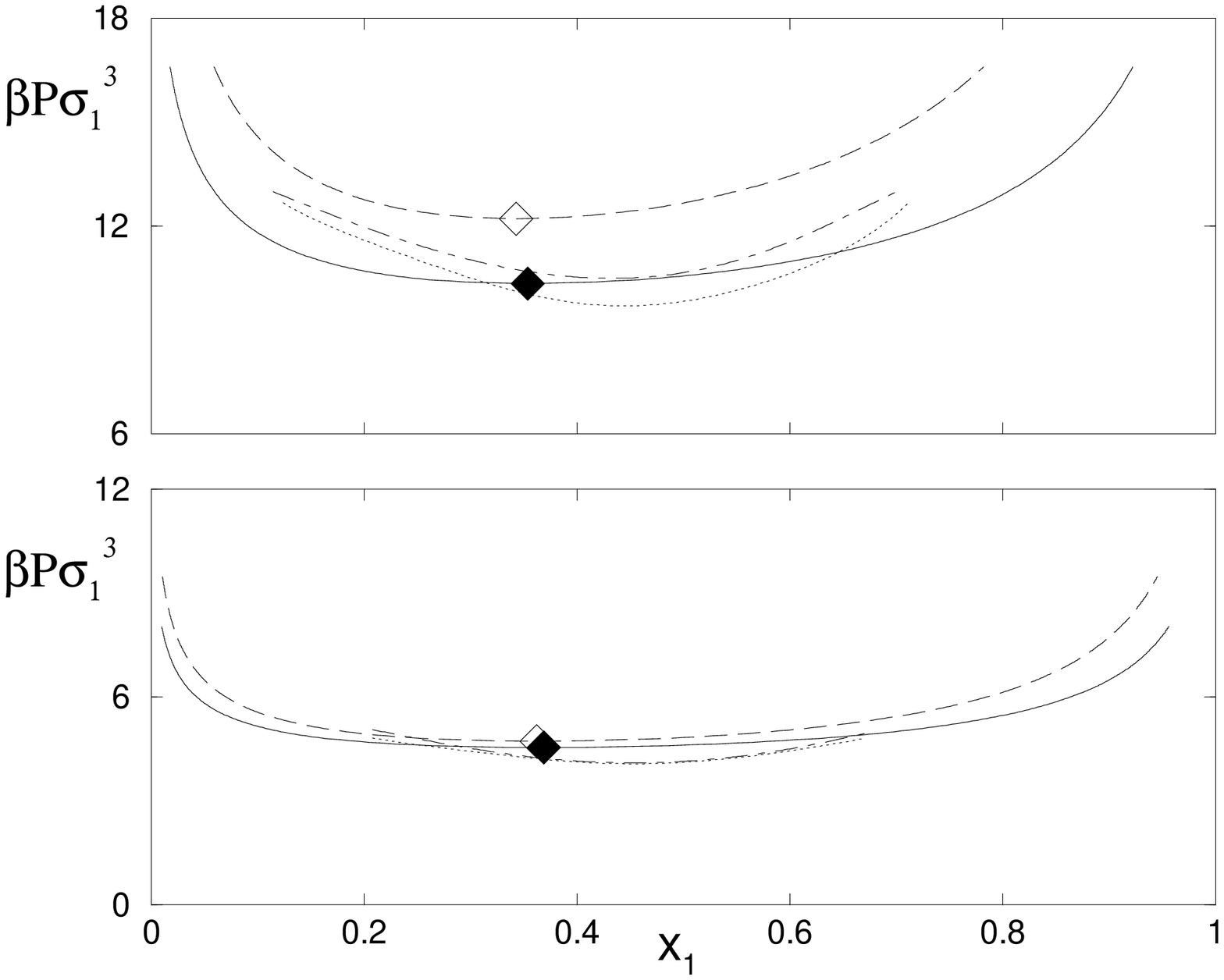}
\end{center}
\caption{Fluid-fluid coexistence in the pressure-composition plane
for $q=0.75$: top panel, $\Delta=0.05$; bottom panel, $\Delta=0.1$. The continuous and dashed curves represent the BG4
and BG3 binodal lines, respectively. The critical point is identified with solid (BG4) and open (BG3) diamonds.
The dotted and dot-dashed lines represent the RY and MHNC predictions, respectively~\cite{pellican06}.} \label{fig4}
\end{figure}

\begin{figure}
\begin{center}
\includegraphics[width=11cm,angle=0]{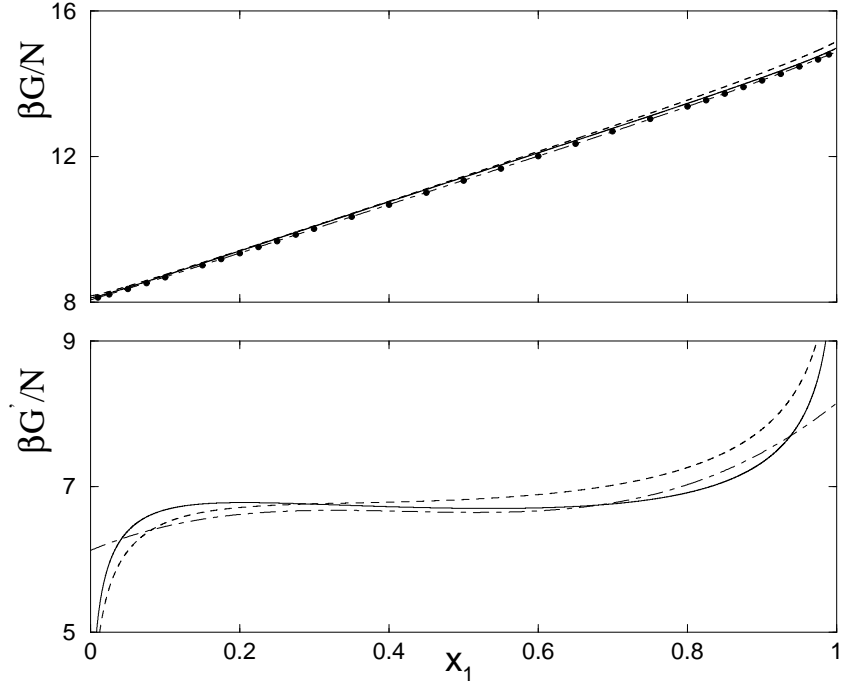}
\end{center}
\caption{Total Gibbs free energy (top panel) and its first
derivative (bottom panel) plotted as a function of the mole fraction for
$q=0.75$, $\Delta=0.05$ and $\beta P \sigma_{1}^{3} = 11.38$.
Dashed line: BG3; continuous line: BG4; solid circles: RY closure;
dot-dashed line: fourth-order polynomial fit of the RY data.}
\label{fig5}
\end{figure}

\begin{figure}
\begin{center}
\includegraphics[width=11cm,angle=0]{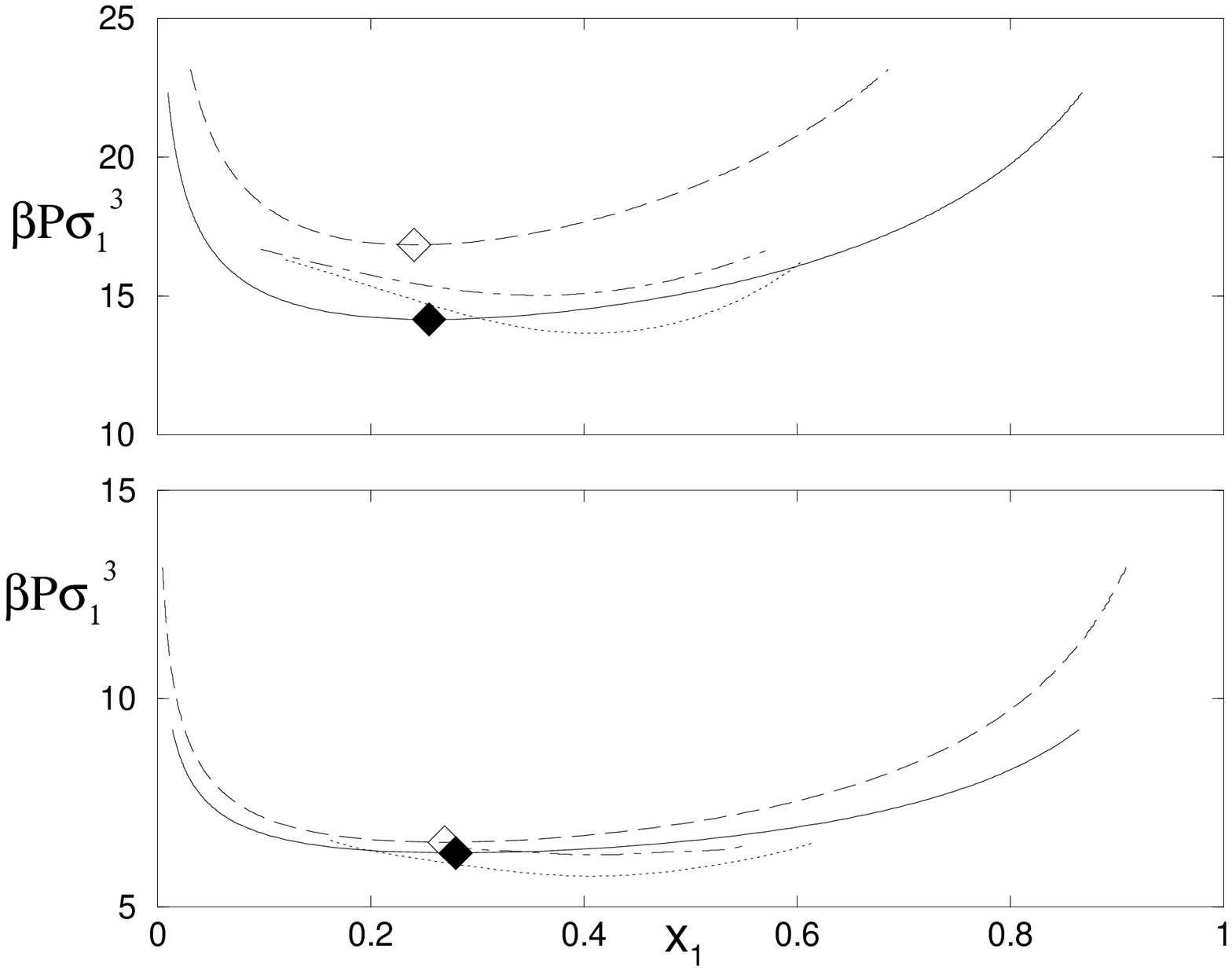}
\end{center}
\caption{Fluid-fluid coexistence in the pressure-composition plane
for $q=0.6$: top panel, $\Delta=0.05$; bottom panel, $\Delta=0.1$. The continuous and dashed curves represent the BG4
and BG3 binodal lines, respectively. The critical point is identified with solid (BG4) and open (BG3) diamonds.
The dotted and dot-dashed lines represent the RY and MHNC predictions, respectively~\cite{pellican06}..} \label{fig6}
\end{figure}

We also show the behavior of the analytical EOS for size ratios
$q=0.75$ (Fig.~\ref{fig4}) and $q=0.6$ (Fig.~\ref{fig5}) and for two
values of the nonadditivity parameter $\Delta=0.05, 0.1$ which are
particularly meaningful for Helium-Xenon mixtures at high
pressure~\cite{pellican06}. In this case the critical point falls at
medium-high packing fractions and we found no simulation data in the
literature to compare with. Hence, we compare the BG3 and BG4
estimates of the phase separation curves with the predictions of two
integral-equation theories~\cite{pellican07}, {\it i.e.}, the
modified hypernetted-chain (MHNC)~\cite{mhnc} and the Rogers-Young
(RY)~\cite{ry} closures of the Ornstein-Zernike equations. In these
approximate theories two internal parameters are adjusted so as to
ensure the equality of the two osmotic compressibilities that are
respectively obtained via number fluctuations or upon
differentiating the virial pressure~\cite{pellican06}. We show in
Fig.~\ref{fig4} (upper panel) the binodal line for $q=0.75$ and
$\Delta = 0.05$. The current numerical estimates for the critical
mole fraction and for the reduced pressure are reported in
Table~\ref{t7}. It is evident that the BG4 critical pressure is in
better agreement with the corresponding integral-equation estimates.
The differences between the BG3 and BG4 expansions are smaller for
$\Delta=0.1$ (see Fig.~\ref{fig4}, bottom panel). Even in this case,
the BG4 critical parameters are in better agreement with the
corresponding MHNC and RY estimates. As for the shape of these
curves, we note that the values of the constant-pressure Gibbs free
energy obtained from integral-equation theories was fitted through a
fourth-order polynomial curve (see Fig.~\ref{fig6}, upper
panel)~\cite{pellican06}. Overall, the resulting agreement with the
current results for the Gibbs free energy is good. However, the
implementation of the Maxwell construction requires the
concentration derivative of the Gibbs free energy (see Fig.~\ref{fig6}).
 The discrepancies observed in the shapes of the
phase-separation curves obtained with the two methods discussed
above (EOS expansions and integral-equation theories) may be likely
 due to the estimate of this latter quantity.
  In fact, we suspect that the derivative of the polynomial
curve used to fit the integral-equation data for the Gibbs free
energy probably is not flexible enough as one is lead to argue after
observing the shapes of the corresponding BG3 and BG4 curves.

The BG4 fluid-fluid critical point is in better agreement with the
corresponding integral-equation estimate also for $q=0.6$ (see
Fig.~\ref{fig5}). In fact, for $\Delta=0.05$ the BG4 critical
pressure, that is reported in Table~\ref{t8}, falls in between the
MHNC and RY values as for $q=0.6$ (see Table~\ref{t7}). As far as
the critical concentration is concerned, in passing from $q=0.75$ to
$q=0.6$, the BG4 estimate moves to lower $x_1$ values in a more
distinct way than its MHNC and RY courterparts. Finally, for larger
values of the nonadditivity parameter ($\Delta=0.1$), the
differences in the critical parameters estimated with these
approaches smooth over, as previously shown for $q=0.75$ (see
Table~\ref{t8}).

\subsection {Highly asymmetric regime}
\begin{figure}
\begin{center}
\includegraphics[width=11cm,angle=0]{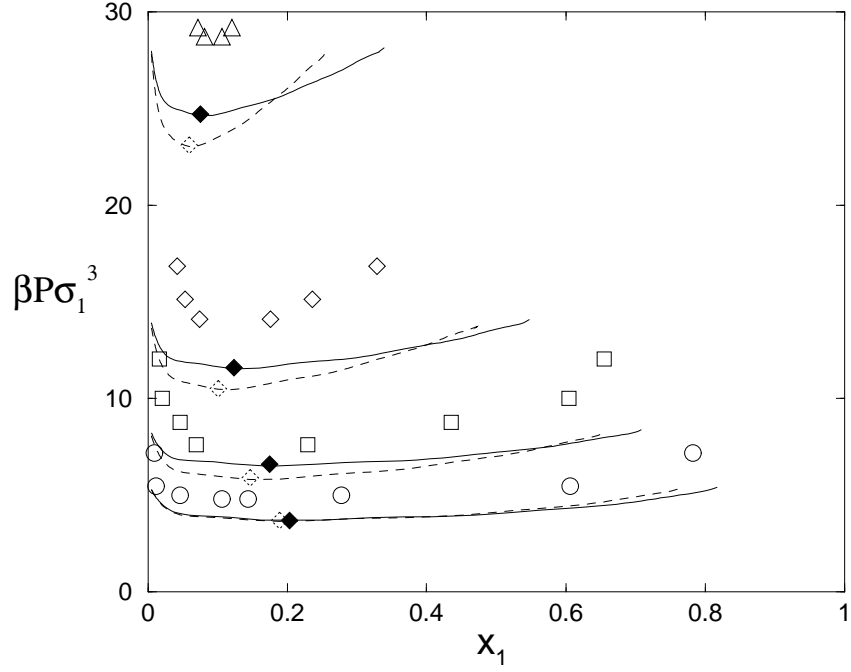}
\end{center}
\caption{Fluid-fluid coexistence in the pressure-composition plane for $q=0.1$. The continuous and dashed
curves are the BG4 and BG3 binodal lines, respectively. The critical point is identified with solid (BG4) and open
(BG3) diamonds. The open symbols represent the GEMC simulation data~\cite{dij2} obtained for $\Delta = 0.5$ (circles),
$0.4$ (squares), $0.3$ (diamonds), $0.2$ (triangles).}
\label{fig7}
\end{figure}

\begin{figure}
\begin{center}
\includegraphics[width=11cm,angle=0]{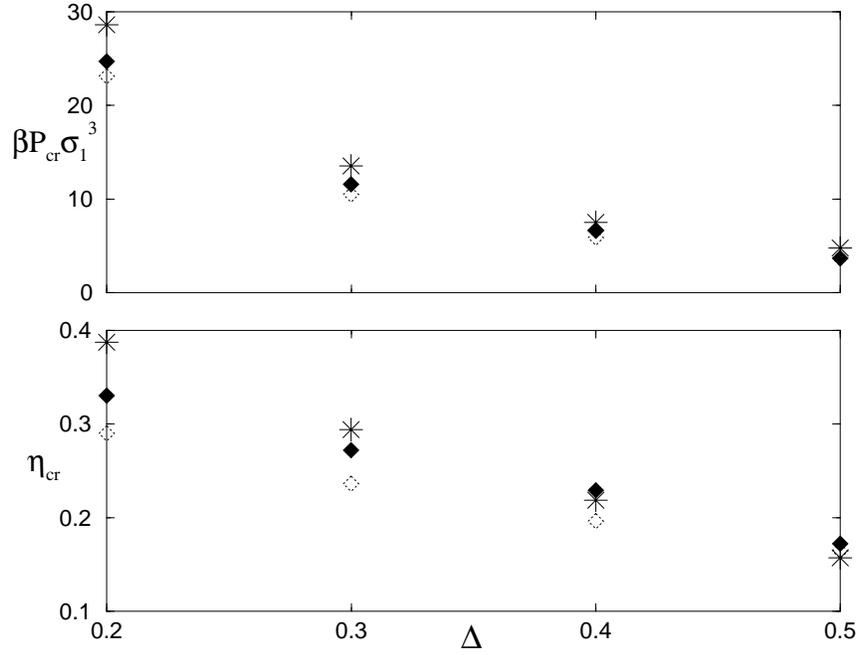}
\end{center}
\caption{Reduced pressure (top panel) and total packing fraction
(bottom panel), calculated at the critical point for a nonadditive hard-sphere
mixture with size ratio $q=0.1$, plotted as a function of the nonadditivity parameter
$\Delta$: open diamonds, BG3; solid diamonds, BG4; stars, simulation
data~\cite{dij2}.}
\label{fig8}
\end{figure}

We start commenting the results obtained for $q=0.1$ and
nonadditivity $\Delta=0.2,0.3,0.4,0.5$. We compare in
Fig.~\ref{fig7} the fluid-fluid coexistence curves obtained by
means of the BG3 and BG4 estiamtes of the EOS with the GEMC
simulations reported in~\cite{dij2}. As already observed for
larger size ratios, the BG4 curve shows a better agreement with
the simulation data in the region around the critical point, and
this improvement is more evident for low values of the
nonadditivity parameter. On the other side, for larger values of
$\Delta$, the BG3 and BG4 estimates become very close to each
other, as was also the case of mixtures with not too dissimilar
species. This effect is probably related to the comparatively
lower packing fractions of the demixing region that is observed
for larger nonadditivities. In fact, the mixture is less
correlated at low densities and one may then expect that the
differences between the fourth and the third-order estimates of
the EOS tend to disappear. Figure~\ref{fig8} shows the critical
pressure and packing fraction plotted as a function of the
nonadditivity parameter. We also observed a crossover in the BG4
critical pressure versus its BG3 value upon further decreasing
$\Delta$: {\it viz.}, the BG4 values get lower than the BG3 ones.
This observation is particularly significant because the BG3
critical pressure tends to diverge in the limit of additive
mixtures ($\Delta=0$), whilst the BG4 value apparently remains
finite.

\begin{figure}
\begin{center}
\includegraphics[width=9cm,angle=0]{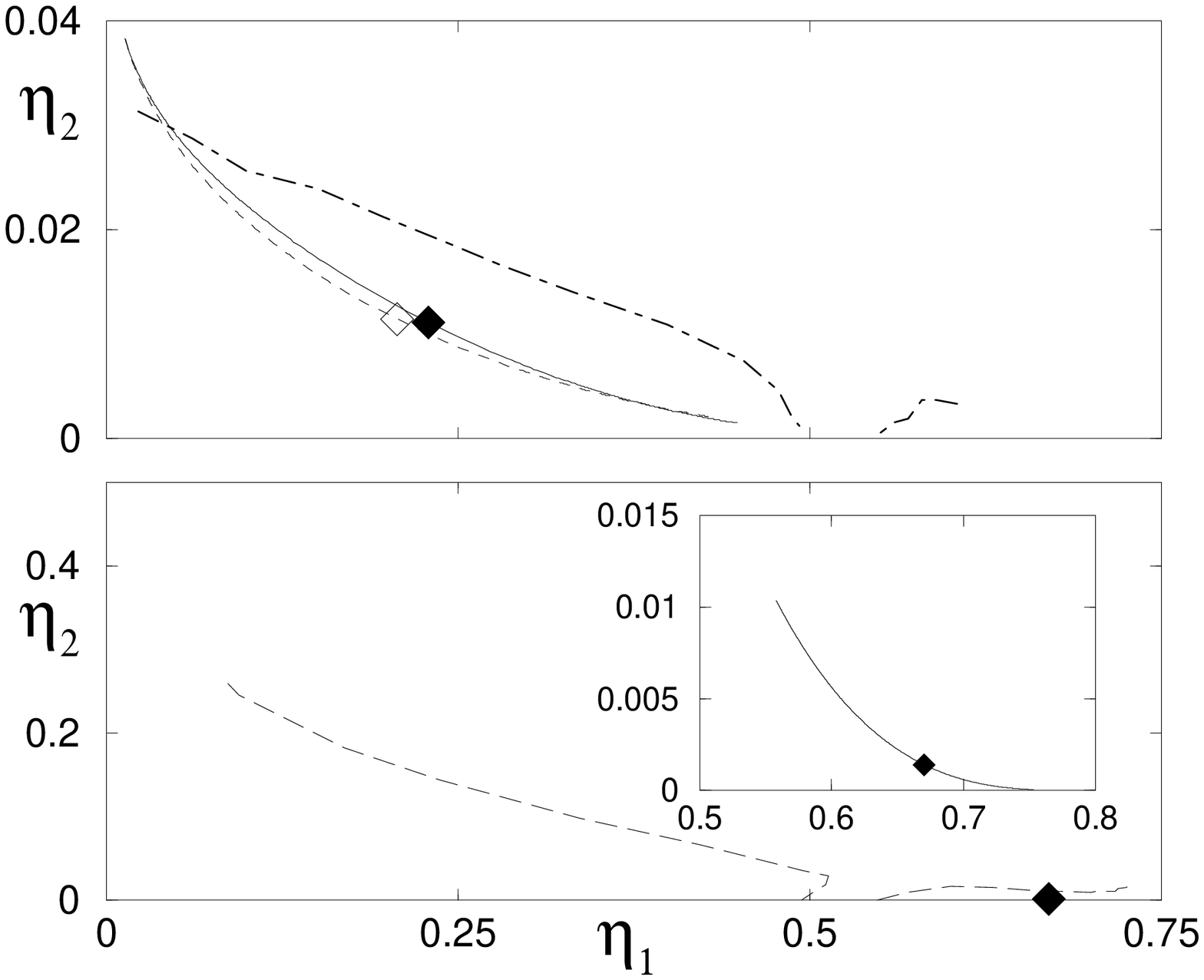}
\end{center}
\caption
{Phase diagram calculated for $q=0.2$ and $\Delta=0.25$ (top panel) and for $q=0.1075$ and
$\Delta=-0.01$ (bottom panel).  Dot-dashed curves: fluid-solid binodal line from~\cite{lou2}; the continuous and dashed
curves represent the BG4 and BG3 binodal lines, respectively. The critical point is identified with solid (BG4) and
open (BG3) diamonds. The continuous lines plotted in the inset represents the BG4 fluid-fluid binodal line. The
orresponding fourth-order partial virial coefficients are: D1112=$7.98(1)\cdot10^{-2}$, D1122=$4.840(4)\cdot10^{-4}$,
D1222=$1.8029(7)\cdot10^{-6}$.}
\label{fig9}.
\end{figure}

As pointed out by many authors, the regime of low or even negative
values of the nonadditivity parameter is interesting for the
application of the model to sterically and electrostatically
stabilized colloidal mixtures. In particular, Louis and coworkers
used the Alexander-de Gennes theory with the Derjaguin
approximation in order to estimate the value of the nonadditivity
parameter for a polymethylmetrylacrilate mixture stabilized by a
poly-12-hydroxysteric acid brush~\cite{lou2}. They obtained
$\Delta=-0.01$ for $q=0.1075$. We present in the bottom panel of
Fig.~\ref{fig9} the BG4 fluid-fluid binodal line with its critical
point for a mixture with size ratio $q=0.1$ and compare it with
the fluid-solid binodal line as reported in~\cite{lou2}. For
negative values of $\Delta$, the demixing transition should be
preempted by the formation of the solid phase because a stable
demixing is expected to occur in very asymmetric mixtures
only~\cite{louis1,santos1}. In ref.~\cite{lou2}, the adopted
crystal structure in order to evaluate the fluid-solid binodal, is
an FCC lattice of large spheres (permeated by a fluid of small
spheres). Actually, we find that demixing occurs for very high
values of the larger-particles packing fraction (see the inset of
Fig.~\ref{fig9}) and for very low values of the smaller-particles
packing fraction. The corresponding BG4 critical point, with
coordinates $\eta_1=0.67$ and $\eta_2=0.0014$, falls well inside
the region of stability of the solid phase. Louis and coworkers
also estimated that for positive values of the nonadditivity
parameter the critical point remains metastable up to $\Delta
\approx 0.2$~\cite{lou2}. In agreement with their prediction, for
$\Delta =0.25$ the fluid-fluid critical point, located by both the
BG3 and BG4 expansions, falls below the freezing line (see the top
panel of Fig.~\ref{fig9}).

\section {Concluding remarks} In this paper we discussed the fourth virial coefficient of binary hard-sphere mixtures,
which has been computed for different values of the nonadditivity parameter and of the size ratio. We have shown that
its use within the $y$ expansion originally proposed by Barboy and Gelbart improves over the description of the
fluid-fluid phase separation provided by the same expansion truncated at the third order, the more so for small values
of the size ratio and of the nonadditivity parameter. In particular, we found that the pressure calculated at the
critical point through the fourth-order expansion does not diverge in the limit of small values of the nonadditivity
parameter and of largely asymmetric size ratios. This feature allowed us to predict that, within this approximation, the
critical point is unstable with respect to crystallization in the regime relevant to colloidal mixtures.

\begin{table}
\caption{Fourth-order partial virial coefficients as a function of
the size ratio for $\Delta =0.05$. The numerical values are given
in units of $\sigma_1^9$. The error on the last significant figure
is enclosed in parentheses.}
\bigskip
\begin {tabular} {|c|l|l|l|}
\hline
 $q$ & $D_{1112}$ & $D_{1122}$ & $D_{1222}$ \\
\hline
 0.05 & 0.0835(3) & 5.53(1)$\cdot10^{-5}$  & 2.064(1)$\cdot10^{-8}$   \\
 0.10 & 0.1201(4) & 5.271(9)$\cdot10^{-4}$ & 1.3981(6)$\cdot10^{-6}$  \\
 0.15 & 0.1661(5) & 2.073(4)$\cdot10^{-3}$ & 1.6840(8)$\cdot10^{-5}$  \\
 0.20 & 0.2225(6) & 5.67(1)$\cdot10^{-3}$  & 9.988(6)$\cdot10^{-5}$   \\
 0.25 & 0.2900(7) & 1.267(2)$\cdot10^{-2}$ & 4.019(2)$\cdot10^{-4}$   \\
 0.30 & 0.3697(8) & 2.485(4)$\cdot10^{-2}$ & 1.2645(8)$\cdot10^{-3}$  \\
 0.40 & 0.569(1)  & 7.46(1)$\cdot10^{-2}$  & 7.857(6)$\cdot10^{-3}$   \\
 0.50 & 0.828(1)  & 0.1806(4)              & 3.303(3)$\cdot10^{-2}$   \\
 0.60 & 1.153(2)  & 0.3807(8)              & 0.1084(1)                \\
 0.70 & 1.554(2)  & 0.728(1)               & 0.2990(3)                \\
 0.75 & 1.784(2)  & 0.977(1)               & 0.4726(3)                \\
 0.80 & 2.036(3)  & 1.292(2)               & 0.7269(8)                \\
 0.85 & 2.309(2)  & 1.683(2)               & 1.0914(9)                \\
 0.90 & 2.606(3)  & 2.166(4)               & 1.604(2)                 \\
 0.95 & 2.927(4)  & 2.755(6)               & 2.311(3)                 \\
\hline
\end {tabular}
\label {t1}
\end{table}
\newpage
\begin{table}
\caption{Fourth-order partial virial coefficients as a function of
the size ratio for $\Delta=0.1$ (see also the caption of Table I).}
\bigskip
\begin {tabular} {|c|l|l|l|}
\hline
 $q$ & $D_{1112}$ & $D_{1122}$ & $D_{1222}$ \\
\hline
 0.05 & 0.1201(4) & 5.37(1)$\cdot10^{-5}$  & 2.3824(8)$\cdot10^{-8}$  \\
 0.10 & 0.1686(7) & 6.15(2)$\cdot10^{-4}$  & 1.620(1)$\cdot10^{-6}$   \\
 0.15 & 0.2284(6) & 2.505(4)$\cdot10^{-3}$ & 1.959(1)$\cdot10^{-5}$   \\
 0.20 & 0.3007(7) & 6.94(1)$\cdot10^{-3}$  & 1.1664(6)$\cdot10^{-4}$  \\
 0.25 & 0.3863(8) & 1.560(2)$\cdot10^{-2}$ & 4.710(3)$\cdot10^{-4}$   \\
 0.30 & 0.4865(9  & 3.073(6)$\cdot10^{-2}$ & 1.486(1)$\cdot10^{-3}$   \\
 0.40 & 0.735(1)  & 9.26(2)$\cdot10^{-2}$  & 9.298(7)$\cdot10^{-3}$   \\
 0.50 & 1.053(2)  & 0.2249(4)              & 3.934(3)$\cdot10^{-2}$   \\
 0.60 & 1.451(2)  & 0.4750(8)              & 0.1298(1)                \\
 0.70 & 1.938(3)  & 0.909(2)               & 0.3602(3)                \\
 0.75 & 2.216(2)  & 1.220(2)               & 0.5710(4)                \\
 0.80 & 2.519(3)  & 1.613(3)               & 0.8810(9)                \\
 0.85 & 2.849(2)  & 2.103(2)               & 1.326(1)                 \\
 0.90 & 3.206(4)  & 2.704(5)               & 1.953(2)                 \\
 0.95 & 3.591(4)  & 3.442(6)               & 2.822(3)                 \\
\hline
\end {tabular}
\label {t2}
\end{table}
\newpage
\begin{table}
\caption{Fourth-order partial virial coefficients as a function of
the size ratio for $\Delta=0.2$ (see also the caption of Table I).}
\bigskip
\begin {tabular} {|c|l|l|l|}
\hline
 $q$ & $D_{1112}$ & $D_{1122}$ & $D_{1222}$ \\
\hline
 0.05 & 0.2226(6) & -2.277(2)$\cdot10^{-4}$ & 3.116(1)$\cdot10^{-8}$  \\
 0.10 & 0.3007(7) &  1.15(2)$\cdot10^{-4}$  & 2.134(1)$\cdot10^{-6}$  \\
 0.15 & 0.3945(8) &  2.022(6)$\cdot10^{-3}$ & 2.595(1)$\cdot10^{-5}$  \\
 0.20 & 0.506(1)  &  7.10(2)$\cdot10^{-3}$  & 1.5554(8)$\cdot10^{-4}$   \\
 0.25 & 0.637(1)  & 1.758(4)$\cdot10^{-2}$ & 6.320(4)$\cdot10^{-4}$   \\
 0.30 & 0.787(1)  & 3.647(7)$\cdot10^{-2}$ & 2.006(1)$\cdot10^{-3}$   \\
 0.40 & 1.154(2)  & 0.1150(2)              & 1.2694(9)$\cdot10^{-2}$  \\
 0.50 & 1.617(2)  & 0.2879(5)              & 5.429(4)$\cdot10^{-2}$   \\
 0.60 & 2.190(3)  & 0.616(1)               & 0.1809(2)                \\
 0.70 & 2.881(4)  & 1.188(2)               & 0.5068(5)                \\
 0.75 & 3.274(3)  & 1.600(2)               & 0.8071(5)                \\
 0.80 & 3.702(4)  & 2.118(4)               & 1.250(1)                 \\
 0.85 & 4.166(3)  & 2.765(4)               & 1.890(1)                 \\
 0.90 & 4.664(5)  & 3.562(6)               & 2.795(3)                 \\
 0.95 & 5.204(6)  & 4.528(8)               & 4.056(4)                 \\
\hline
\end {tabular}
\label {t3}
\end{table}
\newpage
\begin{table}
\caption{Fourth-order partial virial coefficients as a function of
the size ratio for $\Delta=0.3$ (see also the caption of Table I).}
\bigskip
\begin {tabular} {|c|l|l|l|}
\hline
 $q$ & $D_{1112}$ & $D_{1122}$ & $D_{1222}$ \\
\hline
 0.05 & 0.3697(8) & -2.0687(2)$\cdot10^{-4}$ & 3.984(1)$\cdot10^{-8}$  \\
 0.10 & 0.4865(9) & -3.736(3)$\cdot10^{-3}$  & 2.744(1)$\cdot10^{-6}$  \\
 0.15 & 0.625(1)  & -4.857(7)$\cdot10^{-3}$  & 3.357(1)$\cdot10^{-5}$  \\
 0.20 & 0.787(1)  & -3.90(2)$\cdot10^{-3}$   & 2.022(1)$\cdot10^{-4}$  \\
 0.25 & 0.974(2)  & 1.21(5)$\cdot10^{-3}$    & 8.261(5)$\cdot10^{-4}$  \\
 0.30 & 1.188(2)  & 1.339(9)$\cdot10^{-2}$   & 2.635(2)$\cdot10^{-3}$  \\
 0.40 & 1.705(2)  & 7.46(3)$\cdot10^{-2}$    & 1.683(1)$\cdot10^{-2}$  \\
 0.50 & 2.350(3)  & 0.0.2219(7)              & 7.259(5)$\cdot10^{-2}$  \\
 0.60 & 3.139(4)  & 0.514(2)                 & 0.2439(2)               \\
 0.70 & 4.085(5)  & 1.034(3)                 & 0.6886(6)               \\
 0.75 & 4.622(4)  & 1.410(2)                 & 1.1001(7)               \\
 0.80 & 5.203(6)  & 1.188(5)                 & 1.711(2)                \\
 0.85 & 5.833(4)  & 2.477(4)                 & 2.594(2)                \\
 0.90 & 6.506(7)  & 3.204(9)                 & 3.849(4)                \\
 0.95 & 7.233(7)  & 4.09(1)                  & 5.602(5)                \\
\hline
\end {tabular}
\label {t4}
\end{table}
\newpage
\begin{table}
\caption{Fourth-order partial virial coefficients as a function of
the size ratio for $\Delta=0.4$ (see also the caption of Table I).}
\bigskip
\begin {tabular} {|c|l|l|l|}
\hline
 $q$ & $D_{1112}$ & $D_{1122}$ & $D_{1222}$ \\
\hline
 0.05 & 0.569(1)  & -9.0438(3)$\cdot10^{-3}$ & 5.004(1)$\cdot10^{-8}$  \\
 0.10 & 0.734(1)  & -1.7721(2)$\cdot10^{-2}$ & 3.462(2)$\cdot10^{-6}$  \\
 0.15 & 0.929(1)  & -3.002(7)$\cdot10^{-2}$  & 4.256(4)$\cdot10^{-5}$  \\
 0.20 & 1.153(2)  & -4.597(3)$\cdot10^{-2}$  & 2.575(1)$\cdot10^{-4}$  \\
 0.25 & 1.412(2)  & -6.544(6)$\cdot10^{-2}$  & 1.0564(6)$\cdot10^{-3}$ \\
 0.30 & 1.705(2)  & -8.75(1)$\cdot10^{-2}$   & 3.383(2)$\cdot10^{-3}$  \\
 0.40 & 2.406(3)  & -0.1377(4)               & 2.178(1)$\cdot10^{-2}$  \\
 0.50 & 3.274(4)  & -0.1898(9)               & 9.461(7)$\cdot10^{-2}$  \\
 0.60 & 4.327(5)  & -0.236(2)                & 0.3199(2)               \\
 0.70 & 5.584(6)  & -0.280(4)                & 0.9088(7)               \\
 0.75 & 6.295(5)  & -0.299(3)                & 1.4564(9)               \\
 0.80 & 7.062(7)  & -0.314(6)                & 2.270(2)                \\
 0.85 & 7.890(6)  & -0.346(6)                & 3.452(2)                \\
 0.90 & 8.782(9)  & -0.39(1)                 & 5.137(5)                \\
 0.95 & 9.729(9)  & -0.46(1)                 & 7.498(7)                \\
\hline
\end {tabular}
\label {t5}
\end{table}
\newpage
\begin{table}
\caption{Fourth-order partial virial coefficients as a function of
the size ratio for $\Delta=0.5$. See also caption of table I.}
\bigskip
\begin {tabular} {|c|l|l|l|}
\hline
 $q$ & $D_{1112}$ & $D_{1122}$ & $D_{1222}$ \\
\hline
 0.05 & 0.828(1)  & -2.91908(3)$\cdot10^{-2}$ & 6.180(1)$\cdot10^{-8}$  \\
 0.10 & 1.053(2)  & -5.6318(3)$\cdot10^{-3}$  & 4.295(2)$\cdot10^{-6}$  \\
 0.15 & 1.316(2)  & -9.806(1)$\cdot10^{-2}$   & 5.303(4)$\cdot10^{-5}$  \\
 0.20 & 1.617(2)  & -0.15883(3)               & 3.220(2)$\cdot10^{-4}$  \\
 0.25 & 1.962(3)  & -0.24429(7)               & 1.3258(7)$\cdot10^{-3}$ \\
 0.30 & 2.350(3)  & -0.3605(1)                & 4.263(2)$\cdot10^{-3}$  \\
 0.40 & 3.275(4)  & -0.7246(5)                & 2.761(2)$\cdot10^{-2}$  \\
 0.50 & 4.410(5)  & -1.346(1)                 & 0.12067(8)              \\
 0.60 & 5.781(6)  & -2.378(2)                 & 0.4103(3)               \\
 0.70 & 7.409(7)  & -4.064(4)                 & 1.1717(9)               \\
 0.75 & 8.326(6)  & -5.245(4)                 & 1.882(1)                \\
 0.80 & 10.314(9) & -6.718(7)                 & 2.941(2)                \\
 0.85 & 10.379(7) & -8.578(7)                 & 4.482(3)                \\
 0.90 & 11.52(1)  & -10.88(1)                 & 6.686(6)                \\
 0.95 & 12.74(1)  & -13.73(2)                 & 9.777(8)                \\
\hline
\end {tabular}
\label {t6}
\end{table}

\begin{table}
\caption{Critical molar fractions and reduced pressures for $q=0.75$.}
\begin{tabular}{lccccc} \\
\cline{3-6}
$\Delta$ & & & $x_{1}$ & $\beta P \sigma_1^3$  &  \\
\hline
0.05  & BG3  & & 0.34  & 12.21 &     \\
      & BG4  & & 0.36  & 10.34 &     \\
      & MHNC & & 0.44  & 9.72  &     \\
      & RY   & & 0.43  & 10.5  &     \\
      &      & &       &       &     \\
0.1   & BG3  & &  0.36 & 4.73  &     \\
      & BG4  & &  0.37 & 4.54  &     \\
      & MHNC & &  0.455& 4.1   &     \\
      & RY   & &  0.45 & 4.12  &     \\
\hline
\label{t7}
\end{tabular}
\end{table}

\begin{table}
\caption{Critical molar fractions and reduced pressures for $q=0.60$.}
\begin{tabular}{lccccc} \\
\cline{3-6}
$\Delta$ & & & $x_{1}$ & $\beta P \sigma_1^3$  &  \\
\hline
0.05  & BG3  & & 0.241 & 16.85 &     \\
      & BG4  & & 0.255 & 14.15 &     \\
      & MHNC & & 0.405 & 13.7  &     \\
      & RY   & & 0.36  & 15.05 &     \\
      &      & &       &       &     \\
0.1   & BG3  & &  0.27 & 6.55  &     \\
      & BG4  & &  0.28 & 6.30  &     \\
      & MHNC & &  0.41 & 5.74  &     \\
      & RY   & &  0.41 & 6.25  &     \\
\hline
\label{t8}
\end{tabular}
\end{table}

\newpage
\begin{thebibliography}{99}

\bibitem{bou-nez86} Boubl{\'\i }k, T.; Nezbeda, I.
         {\it Collect. Czech. Chem. Commun.} {\bf 1986}, {\it 51}, 2301.

\bibitem{rowlinson} Rowlinson, J. S.
         {\it Liquids and liquid mixtures} (Butterworth, London, {\bf 1969}).

\bibitem{shout} Shouten, J. A.
         {\it Int J. of Thermophys.} {\bf 2001}, {\it 22}, 23.

\bibitem{louis} Louis, A. A.
         {\it Philos. T. Roy. Soc. A} {\bf 2001}, {\it 359}, 939.

\bibitem{frenk} Frenkel, D.
         {\it J. Phys.: Condens. Matter} {\bf 1994}, {\it 6}, A71-A78.

\bibitem{bh1} Biben, T.; Hansen, J. P.
         {\it Phys. Rev. Lett.} {\bf 1991}, {\it 66}, 2215.

\bibitem{dre} Dijkstra, M.; van Roij, R.; Evans, R.
         {\it Phys. Rev. Lett.} {\bf 1998}, {\it 81}, 2268;
         {\it ibid.} {\bf 1999}, {\it 82}, 117;
         {\it Phys. Rev. E} {\bf 1999}, {\it 59}, 5744.

\bibitem{albas} Albas, P.; van der Marel, C.; Geertsman, W.; Meijer, J. A.; van
         Osten, A. B.; Dijkstra, J.; Stein, P. C.; van der Lugt, W.
         {\it J. non-crystalline Solids} {\bf 1984}, {\it 61/62}, 201.

\bibitem{gaz-past1} Gazzillo, D.; Pastore, G.; Enzo, S.
         {\it J. Phys.: Condens. Matter} {\bf 1989}, {\it 1}, 3469.

\bibitem{gaz-past2} Gazzillo, D.; Pastore, G.; Frattini, R.
         {\it J. Phys.: Condens. Matter} {\bf 1990}, {\it 2}, 8463.

\bibitem{tenne} Tenne, R.; Bergmann, E.
         {\it Phys. Rev. A} {\bf 1978}, {\it 17}, 2036.

\bibitem{mazo} Mazo, R.; Bearman, R. J.
         {\it J. Chem. Phys.} {\bf 1990}, {\it 93}, 6694.

\bibitem{nix} Nixon, J. H.; Silbet, M.
         {\it Molec. Phys.} {\bf 1984}, {\it 52}, 207.

\bibitem{bpgg} Ballone, P.; Pastore, G.; Galli, G.; Gazzillo, D.
         {\it Molec. Phys.} {\bf 1986}, {\it 59}, 275.

\bibitem{gazzillo} Gazzillo, D.
         {\it J. Chem. Phys.} {\bf 1991}, {\it 95}, 4565.

\bibitem{lala} Lomba, E.; Alvarez, M.; Lee, L. L.; Almarza, N. E.
         {\it J. Chem. Phys.} {\bf 1996}, {\it 104}, 4180.

\bibitem{kahl} Kahl, G.; Bildstein, B.; Rosenfeld, Y.
         {\it Phys. Rev. E} {\bf 1996}, {\it 54}, 5391.

\bibitem{sfg} Saija, F.; Fiumara, G.; Giaquinta, P. V.
         {\it J. Chem. Phys.} {\bf 1998}, {\it 108}, 9098.

\bibitem{spg} Saija, F.; Pastore, G.; Giaquinta, P. V.
         {\it J. Phys. Chem. B} {\bf 1998}, {\it 102}, 10368.

\bibitem{ree2} Jung, J.; Jhon, M. S.; Ree, F. H.
         {\it J. Chem. Phys.} {\bf 1995}, {\it 102}, 1349.

\bibitem{hamad1} Hamad, E. Z.
         {\it J. Chem. Phys.} {\bf 1996}, {\it 105}, 3222.

\bibitem{hamad2} Hammawa, H.; Hamad, E. Z.
         {\it J. Chem. Soc., Faraday Trans.} {\bf 1996}, {\it 92}, 4943.

\bibitem{melynk} Melnyk, T. W.; Sawford, B. L.
         {\it Molec. Phys.} {\bf 1975}, {\it 29}, 891.

\bibitem{adams} Adams, D. J.; McDonald, I. R.
         {\it J. Chem. Phys.} {\bf 1975}, {\it 63}, 1900.

\bibitem{gapa} Gazzillo D.; Pastore G.
         Chem. Phys. Lett. (1989) 159, 388.

\bibitem{amar} Amar, J. G.;
         {\it Molec. Phys.} {\bf 1989}, {\it 67}, 739.

\bibitem{yethi} Jagannathan, K; Yethiraj, A.
         {\it J. Chem. Phys.} {\bf 2003}, {\it 118}, 7907.

\bibitem{goz} G{\'\o }zdz, W. T.
         J. Chem. Phys. 119, (2003), 3309.

\bibitem{rovere} Rovere, M.; Pastore, G.
         {\it J. Phys.: Condens. Matter} {\bf 1994}, {\it 6}, A163.

\bibitem{hamad3} Hamad, E. Z.
         {\it Molec. Phys.} {\bf 1997}, {\it 91}, 371.

\bibitem{dij2} Dijkstra, M.
         {\it Phys. Rev. E} {\bf 1998}, {\it 58}, 7523.

\bibitem{bh2} Biben, T.; Hansen, J.-P.
         {\it Physica A} {\bf 1997}, {\it 235}, 142.

\bibitem{lou2} Louis, A. A.; Finken, R.; Hansen, J.-P.
         {\it Phys. Rev. E} {\bf 2000}, {\it 61}, R1028.

\bibitem{re} Roth, R.; Evans, R.
         {\it Europhys. Lett.} {\bf 2001}, {\it 53}, 271.

\bibitem{lr} Louis, A. A.; Roth, R.,
         {\it J. Phys.: Condens. Matter} {\bf 2001}, {\it 13}, L777.

\bibitem{louis1} Roth, R.; Evans, R.; Louis, A. A.,
         {\it Phys. Rev. E} {\bf 2001}, {\it 64}, 051202.

\bibitem{pellican06} Pellicane, G.; Saija, F.; Caccamo, C.; Giaquinta, P. V.
         {\it J. Phys. Chem. B} {\bf 2006}, {\it 110}, 4359.

\bibitem{sg} Saija, F.; Giaquinta, P. V.
         {\it J. Phys. Chem. B} {\bf 2002}, {\it 106}, 2035.

\bibitem{schaink} Schaink, H. M.
         {\it Z. Naturforsch., A: Phys. Sci.} {\bf 1993}, {\it 48}, 899.

\bibitem{santos} Santos, A.; Lopez de Haro, M.; Yuste, S. B.
         {\it J. Chem. Phys.} {\bf 2005}, {\it 122}, 024514.

\bibitem{vlasov} Vlasov, A. Y.; Masters, A. J.
         {\it Fluid Phase Equilib.} {\bf 2003}, {\it 212}, 183.

\bibitem{barboy1} Barboy, B.; Gelbart, W. N.
         {\it J. Chem. Phys.} {\bf 1979}, {\it 71}, 3053;
         {\it J. Stat. Phys.} {\bf 1980}, {\it 22}, 709.

\bibitem{fiumara} Saija, F.; Fiumara, G.; Giaquinta, P. V.
         {\it Molec. Phys.} {\bf 1996}, {\it 87}, 991.

\bibitem{hansen} Hansen, J. -P., McDonald, I. R.
         {\it Theory of simple liquids} (Academic Press, London, {\bf 1986}).

\bibitem{bijor} Bjorling, M.; Pellicane, G.; Caccamo, C.
         {\it J. Chem. Phys.} {\bf 1999}, {\it 111}, 6884.

\bibitem{pellican07} Pellicane, G.; Saija, F.; Caccamo, C.; Giaquinta, P. V.
         unpublished.

\bibitem{mhnc} Rosenfeld, Y.; Aschroft, N.
         {\it Phys. Rev. A} {\bf 1979}, {\it 20}, 1208.

\bibitem{ry} Rogers, F.; Young, D. A.
         {\it Phys. Rev. A} {\bf 1984}, {\it 30}, 999.

\bibitem{santos1} A. Santos, A.; Lopez de Haro, M.
         {\it Phys. Rev. E} {\bf 2005} {\it 72}, 010501(R).

\end{thebibliography}
\end{document}